# MPd$_5$ kagome superconductors studied by density functional calculations

Dan Li[1], Zhengxuan Wang[1], Panshi Jing[1], Mehrdad Shiri[2], Kun Wang[2,†], Chunlan Ma[3,£], Shijing Gong[4], Chuanxi Zhao[5], Tianxing Wang[1], Xiao Dong[1], Lin Zhuang[6], Wuming Liu[7,§], and Yipeng An[1,8,*]

[1] *School of Physics, Henan Normal University, Xinxiang 453007, China*
[2] *Department of Physics, University of Miami, Coral Gables, Florida 33146, USA*
[3] *School of Physics and Technology, Suzhou University of Science and Technology, Suzhou 215009, China*
[4] *Department of Physics, East China Normal University, Shanghai 200062, China*
[5] *Department of Physics, Siyuan Laboratory, Guangdong Provincial Engineering Technology Research Center of Vacuum Coating Technologies and New Energy Materials, Jinan University, Guangzhou 510632, China*
[6] *State Key Laboratory of Optoelectronic Materials and Technologies, School of Physics, Sun Yat-Sen University, Guangzhou 510275, China*
[7] *Beijing National Laboratory for Condensed Matter Physics, Institute of Physics, Chinese Academy of Sciences, Beijing 100190, China*
[8] *Henan Key Laboratory of Advanced Semiconductor & Functional Device Integration, Henan Normal University, Xinxiang 453007, China*

Kagome materials, which are composed of hexagons tiled with a shared triangle, have inspired enormous interest due to their unique structures and rich physical properties; exploring superconducting material systems with new kagome structures is still an important research direction. Here, we predict a type of kagome superconductor, $M$Pd$_5$ ($M$ is a group-IIA metal element), and identify that it exhibits coexistence of superconductivity and nontrivial topological properties. We uncover its phonon-mediated superconductivity by the density functional theory for superconductors, predicting the superconducting transition temperatures ($T_c$) of 2.64, 2.03, and 1.50 K for CaPd$_5$, SrPd$_5$, and BaPd$_5$, respectively. These $T_c$ can be effectively tuned through the application of external pressure and electron doping. The present results also demonstrate that $M$Pd$_5$ have topological properties; e.g., CaPd$_5$ shows topological nontrivial intersection near the Fermi level ($E_F$). Our results indicate that $M$Pd$_5$ materials can be an emerging material platform with rich exotic physics in their kagome structures, and render themselves excellent candidates for superconducting and advanced functional materials that could be utilized in topological quantum computing and information technology.

## I. INTRODUCTION

In recent years, kagome [1-3] materials have attracted great interest due to their unique geometric structures. So far, several types of kagome materials have been theoretically predicted or prepared in experiments [4-6]. We categorize the primary types of reported kagome materials as follows (see Table I). The kagome layer of (11) type (*e.g.*, CoSn, FeSn) comprises Co/Fe [7-11] metal atoms, while the kagome layer of (13)-type boride crystal structure consists of B atoms with embedded metal atoms, such as MgB$_3$ [12], $M$B$_3$ ($M$ = Be, Ca, Sr) [13], and MnB$_3$ [14]. The kagome layer of (16) type, which includes $M$B$_6$ [15, 16] and LiC$_6$ [17], contains nonmetal atoms. For the (31) type, exemplified by Mn$_3X$ ($X$ = Sn, Ge, Ga) [18, 19], the structure features one breathing-type kagome layer composed of Mn atoms. The typical compound of (38) type is Nb$_3X_8$ ($X$ = Cl, Br, I) [20, 21], which exhibits a trimeric kagome structure. The kagome layer of (53) type, represented by Fe$_5$Sn$_3$ [7], is only composed of Fe atoms. In the typical kagome compounds of (132) type such as LaRu$_3$Si$_2$ [22] and LaIr$_3$Ga$_2$ [23], Ru/Ir atoms form the kagome layer, while Si/Ga atoms form a planar honeycomb structure. The (135)-type structure, as seen in $A$V$_3$Sb$_5$ ($A$ = K, Rb, Cs) [24-30], has kagome layers formed by V atoms. In (166)-type structures like, Y$T_6$Sn$_6$ ($T$ = V, Nb, Ta) [31] and $A$V$_6$Sb$_6$ ($A$ = K, Rb, Cs) [32, 33], the transition-metal atoms define their kagome layer. The (322) type, represented by Co$_3$Sn$_2$S$_2$ [19, 34], contains a composite Co$_3$Sn kagome layer. The (326)-type structures, such as Mn$_3$Si$_2$Te$_6$ [35], feature one layer of Mn atoms (Mn1) with a honeycomb structure and the other layer of Mn atoms (Mn2) in a sparser triangular lattice. Other special types include (12)-type kagome structures (e.g., NbS$_2$ [36], CeRu$_2$ [37]) and (227) type (e.g., Lu$_2$V$_2$O$_7$ [38]). These materials, while hosting the kagome lattice, exhibit diverse structural and electronic characteristics, reflective of differences in both atomic composition and geometric arrangement.

*Contact author: ypan@htu.edu.cn
†Contact author: kxw794@miami.edu
£Contact author: wlxmcl@mail.usts.edu.cn
§Contact author: wliu@iphy.ac.cn

TABLE I. The type of kagome materials and typical structures, as well as their electronic properties (EP) and superconductivity. The EP mainly include paramagnetic metal (PM), antiferromagnetic metal (AFM), ferromagnetic metal (FM), magnetic metal (MM), nonmagnetic metal (NMM), magnetic semi-metal (MSM), ferromagnetic semiconductor (FSC), nonmagnetic semi-metal (NMSM), and ferromagnetic insulator (FMI).

| Type | Typical structures | EP | Superconducting |
|---|---|---|---|
| 11 | CoSn [9-11] | PM | No |
|  | FeSn [7, 8, 10] | AFM | No |
|  | FeGe [39-43] | MM | No |
| 12 | NbS$_2$ [36] | MM | Yes |
|  | CeRu$_2$ [37] | MM | Yes |
|  | RbBi$_2$ [44] | NMM | Yes |
| 13 | MgB$_3$ [12] | NMM | Yes |
|  | MB$_3$ (M = Be, Ca, Sr) [13] | NMM | Yes |
|  | MnB$_3$ [14] | NMM | Yes |
|  | CrI$_3$ [45] | MM | No |
| 16 | TMC$_6$ (TM = Mo, W) [46] | NMM | No |
|  | MB$_6$ (M = Mg, Ca, Sc, Ti) [15, 16] | NMM | Yes |
|  | LiC$_6$ [17] | FM | Yes |
| 31 | Mn$_3$X (X = Sn, Ge, Ga) [18, 19] | AFM | No |
|  | Ni$_3$Sn [47] | NMM | Yes |
|  | Ni$_3$In [48] | FM | Yes |
| 32 | M$_3$Sn$_2$ (M = Fe, Ni, Cu) [49] | MM | Yes |
|  | V$_2$O$_3$ [50] | FM | No |
|  | Na$_3$Te$_2$ [51] | MSN | No |
| 38 | Nb$_3$X$_8$ (X = Cl, Br, I) [20, 21] | FSC | No |
| 53 | Fe$_5$Sn$_3$ [7] | FM | No |
| 132 | LaRu$_3$Si$_2$ [22] | NMM | Yes |
|  | LaIr$_3$Ga$_2$ [23] | MM | Yes |
| 135 | AV$_3$Sb$_5$ (A = K, Rb, Cs) [24-30] | NMM | Yes |
|  | ANb$_3$Bi$_5$ (A = K, Rb, Cs) [52] | NMM | Yes |
| 166 | YT$_6$Sn$_6$ (T = V, Nb, Ta) [31] | NMM | Yes |
|  | AV$_6$Sb$_6$ (A = K, Rb, Cs) [32, 33] | NMSM | No |
|  | RV$_6$Sn$_6$ (R = Gd, Ho) [53] | PM | No |
|  | YMn$_6$Sn$_6$ [54] | AFM | No |
| 227 | Lu$_2$V$_2$O$_7$ [38] | FMI | No |
| 322 | Co$_3$Sn$_2$S$_2$ [19, 34] | FSM | No |
| 326 | Mn$_3$Si$_2$Te$_6$ [35] | FMSC | No |
| 15 | MPd$_5$ (M ∈ IIA group element) | NMM | Yes |

These kagome materials exhibit a rich variety of geometric configurations and possess diverse electronic properties, including distinct superconducting behaviors. The majority of kagome materials exhibit metallic behavior, with only a few classified as semiconductors or insulators. About one-third of these materials are non-magnetic, while the remaining display various magnetic properties, including ferromagnetic, antiferromagnetic, and paramagnetic states. Interestingly, nearly half of the kagome materials are superconductors, providing a fertile ground for exploring unconventional superconductivity. These abundant kagome materials support a variety of peculiar electronic states, such as flat bands (FBs) [55-58], van Hove singularities [59-61], and Dirac fermions [62], which make them an excellent platform for studying exotic superconducting and topological properties. Their rich electronic structures have revealed intriguing behavior in superconductivity [63-65] and nontrivial topological environments, such as spin liquid phase [66, 67], charge-density waves [68-70], unconventional charge order [71], and large anomalous Hall responses [72, 73]. Kagome materials exhibit a wealth of physical properties, and it is of great research significance to discover more kagome materials and regulate their rich material states. For example, the coexistence of topological order and superconductivity provides more opportunities for realizing topological superconductors and investigating the complex interactions between superconductivity and non-trivial topological states.

Here, we introduce a type of kagome superconductors, designated as the (15)-type $M$Pd$_5$ (where $M$ represents a group IIA metal element) materials. Through comprehensive first-principles calculations, we systematically investigate their geometric structure, electronic properties, superconductivity, topological characteristics, and mechanical properties. We explore the superconducting behavior of $M$Pd$_5$ materials and demonstrate how it can be modulated by carrier doping and external pressure. These findings position the $M$Pd$_5$ family of materials as promising kagome platforms, broadening the spectrum of kagome structures and creating new prospects for applications in superconductivity and topology physics.

## II. CALCULATION METHODS

This study employs the QUANTUM ESPRESSO (QE) [74] program for first-principles self-consistent and electron-phonon coupling (EPC) calculations. The SG15 optimized norm conservation Vanderbilt [75-77] pseudopotentials are used to describe the influence of the core electron. The Perdew-Burke-Ernzerhof [78-80] function with generalized gradient approximation is applied to the kagome compounds MPd$_5$. The kinetic energy cutoff for wave functions (charge density and potential) is set to 80 (320) Ry. A 12 × 12 × 6 Monkhorst-Pack k point grid is used for self-consistent calculation, while a denser grid (24 × 24 × 12) is employed for density of states (DOS) and band structure. A coarser grid (6 × 6 × 3) is utilized for phonon calculations to balance computational accuracy and cost. In the geometric structure optimization, the total energy tolerance and residual force on each atom are set to < $10^{-10}$ Ry and $10^{-8}$ Ry bohr$^{-1}$, respectively. The phonon properties are obtained by the density functional perturbation theory [81]. The optimized tetrahedron method [82] is used in the Brillouin-zone integration, allowing us to obtain the EPC interaction.

TABLE II. Lattice constants (L), dynamic stability, convex-hull energy, cumulative EPC intensity $\lambda(\omega)$, and $T_c$ of the $M$Pd$_5$ materials.

| System | L (Å) a/b | L (Å) c | Stability | Convex-hull energy (per atom) | $\lambda(\omega)$ | $T_c$ (K) |
|---|---|---|---|---|---|---|
| BePd$_5$ | 5.04 | 4.60 | No | | | |
| MgPd$_5$ | 5.18 | 4.54 | No | | | |
| CaPd$_5$ | 5.34 | 4.47 | Yes | 0 | 0.414 | 2.64 |
| SrPd$_5$ | 5.45 | 4.44 | Yes | 0 | 0.413 | 2.03 |
| BaPd$_5$ | 5.57 | 4.42 | Yes | 0 | 0.377 | 1.50 |

The topological and nodal superconductivity properties of kagome $M$Pd$_5$ are examined using the recently established symmetry-indicators (SIs) method [83-85]. The topological properties are obtained using the QEIRREPS code [86], based on the self-consistent results from the QE code [74], which has been effectively applied to study topological and nodal superconductors. By observing the rich topological nonminor crossings near the $E_F$ of CaPd$_5$ in the band diagram, we obtain the surface states using the WANNIER 90 [87] and WANNIER TOOLS codes [88]. The symmetric operation matrix and irreducible representation of the material are obtained using irvsp [89]. Phonon-mediated superconductivity calculations are performed using the SCDFT method [90-95] within the SUPERCONDUCTING-TOOLKIT code [90-95]. The influence of external pressure and carrier doping on superconductivities is studied by adjusting different pressure and carrier concentration in phonon calculation. The elastic constants of $M$Pd$_5$ bulk materials are determined by the energy-strain method. The elastic modulus, Poisson's ratio and elastic stability criteria are obtained after pre-processing with the Vienna *Ab initio* Simulation Package code (VASP) [96] and post processing with VASPKIT [97].

The transition temperature $T_c$ is obtained using the bisection method [91, 93, 94, 98]. Namely, the initial lower limit of $T_c$ is set to zero and the initial upper limit is set to the $T$ estimated by the BCS theory ($2\Delta_0/3.54$, where $\Delta_0$ is the superconducting gap averaged over Fermi surfaces at 0 K). Then we repeat the bisection step ten times, and $T_c$ is obtained as the average of 0 and $T$, which are very close to each other. More details can be found in Ref. [91].

### III. RESULTS AND DISCUSSION
#### A. Crystal structures of MPd$_5$ kagome materials

In this paper, we predict the hexagonal kagome $M$Pd$_5$ materials that are nonmagnetic metal (see Table I). Figure 1(a) illustrates their bulk structure, whose space group is $P6/mmm$ (No. 191) with the point group $D_{6h}$. Their lattice constants, dynamical stability, convex-hull energies, as well as other physical properties, are shown in Table II. Additionally, we also perform the calculations of mechanical properties and molecular dynamics simulations of the $M$Pd$_5$ family of materials, which are provided in Figs. S1 to S4 of Supplemental Material [100]. The results domonstrate that CaPd$_5$, SrPd$_5$, and BaPd$_5$ are stable, while BePd$_5$ and MgPd$_5$ are dynamically unstable. The electronic properties near the $E_F$ of $M$Pd$_5$ are predominantly derived from the Pd atoms forming the kagome layer, as illustrated in Fig. 1(b), with the $M$ atoms occupying the central positions within the kagome structure [see Fig. 1(a)]. Note that we also investigated the stability of the pristine Pd$_5$ kagome structure and calculated its phonon spectrum [see Fig. S5 in Supplemental Material [100]]. However, a notable phonon softening phenomenon appears when the center metal atoms are absent. The significant virtual frequency indicates that the pristine Pd$_5$ kagome structure is dynamically unstable.

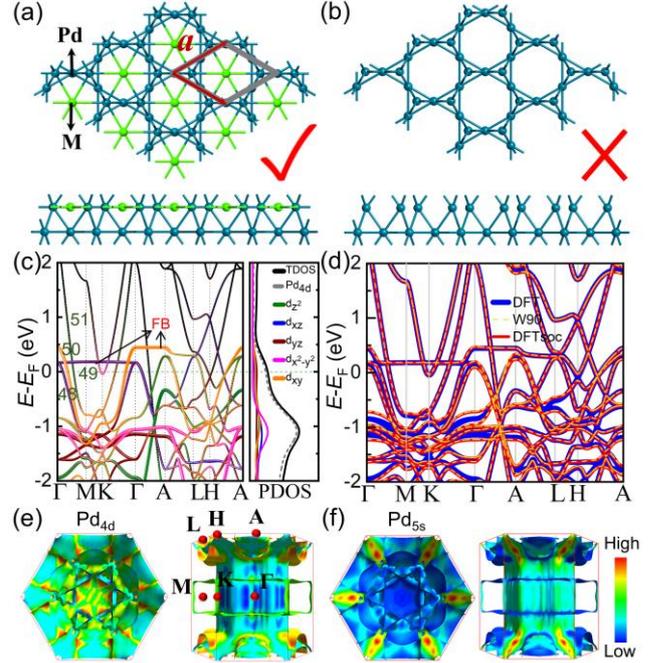

FIG. 1. Top and side view of (a) $M$Pd$_5$ (M is a group-IIA metal element) materials and (b) pristine Pd$_5$ kagome structure. (c) Projected electron band and density of states (DOS) of $d_{z^2}$, $d_{xz}$, $d_{yz}$, $d_{x^2-y^2}$, and $d_{xy}$ orbitals of CaPd$_5$, along with the total density of states (TDOS). The $E_F$ is set to zero. The numbers 48-51 refer to the four bands that cross the $E_F$. (d) Band obtained by density functional thery (DFT) without SOC (DFT) and with SOC (DFTsoc), and by tight-binding model *via* wannier90 code (W90). (e) Top and side views of Fermi velocity $|v_F|$ projected on Fermi surfaces by using the FERMI-SURFER code [99] for (e) Pd$_{4d}$ and (f) Pd$_{5s}$ orbitals of CaPd$_5$. High-symmetry points of the first Brillouin zone of CaPd$_5$ are indicated in (e).

First, we examine the electronic structures of $M$Pd$_5$. A prominent characteristic in the electronic band structures of CaPd$_5$ is the presence of a FB near the $E_F$ along the Γ-M-K-Γ and Γ-A-L paths [see bands 49 and 50 in Fig. 1(c)], leading to a high electronic density of states (DOS) near $E_F$, a feature known to promote superconductivity [9, 101]. According to the element-projected band structure and orbital-projected DOS (PDOS) [see Fig. 1(c)], it is found that the electronic states near the FB and $E_F$ are primarily contributed by the 4d orbital of Pd atoms. Further PDOS analysis reveals that the contributions of $d_{z^2}$, $d_{xz}$, $d_{yz}$, $d_{x^2-y^2}$, and $d_{xy}$ are approximately equal near $E_F$. Similar electronic structures are observed in SrPd$_5$ and BaPd$_5$, as detailed in Figs. S6 and S7 of the Supplemental Material [100]. In addition, the Fermi velocity $|v_F|$ of CaPd$_5$ has a larger projection from Pd$_{4d}$ orbitals [see Fig. 1(e)] than Pd$_{5s}$ orbitals [see Fig. 1(f)] on the Fermi surface, which is defined by four bands intersecting $E_F$ [see bands 48 and 51 in Fig. 1(c)]. Similarly, for SrPd$_5$ and BaPd$_5$, the Pd$_{4d}$ orbitals also exhibit significant contribution to their $|v_F|$ (see Figs. S6 and S7 [100]).

In the first Brillouin zone, $|v_F|$ of all these materials exhibit a significant anisotropy: for CaPd$_5$, a maximum (max) to minimum (min) ratio of 52 for the Pd$_{4d}$ orbitals [see Fig. 1(d)]. The projections of these four bands across $E_F$ are shown in Fig. S8 of the Supplemental Material [100], indicating that the Pd$_{4d}$ orbitals dominate the carrier motion and conductivity of CaPd$_5$. The three steep bands crossing $E_F$ in CaPd$_5$ [see bands 48, 50, and 51 in Fig. 1(c)] make a significant contribution in $|v_F|$. The coexistence of flat and steep bands with localized and ultra-mobile electrons has been shown to enhance superconductivity [64, 102, 103]. This motivates us to further study the superconductivity of the MPd$_5$ materials. Note that the latter topological properties are obtained in the case of spin-orbit coupling (SOC) effects, but this effect is not taken into account in phonon correlation calculations because it leaves the band structure around the $E_F$ unchanged [see Fig. 1(e)], and is less important in describing vibrational and superconducting properties.

**B. Anisotropic electron-phonon coupling**

The remarkable electronic characteristics of the MPd$_5$ kagome materials present a compelling motivation for further exploration of their phonon, electron-phonon coupling (EPC), and phonon-mediated superconductivity. The FBs with localized electrons and steep bands with high Fermi velocities interact differently with phonons [104]. The coexistence of large flat bands and steep bands near $E_F$ is important for the "flat-band/steep-band" scheme of superconductivity [63-65, 103]. For the hexagonal kagome CaPd$_5$ lattice, the FBs along the Γ-M-K-Γ and A-L-H paths, combined with steep bands exhibiting high Fermi velocities, are expected to lead to a higher $T_c$. Analysis of the phonon dispersion (see Fig. 2) confirms the absence of virtual frequencies for CaPd$_5$, SrPd$_5$, and BaPd$_5$. Additionally,

Table II indicates that these three materials are stable at zero pressure. In contrast, BePd$_5$ and MgPd$_5$ display phonon softening phenomena, evidenced by significant virtual frequencies in their phonon spectra (see Fig. S9 in Supplemental Material [100]), indicating their dynamic instability.

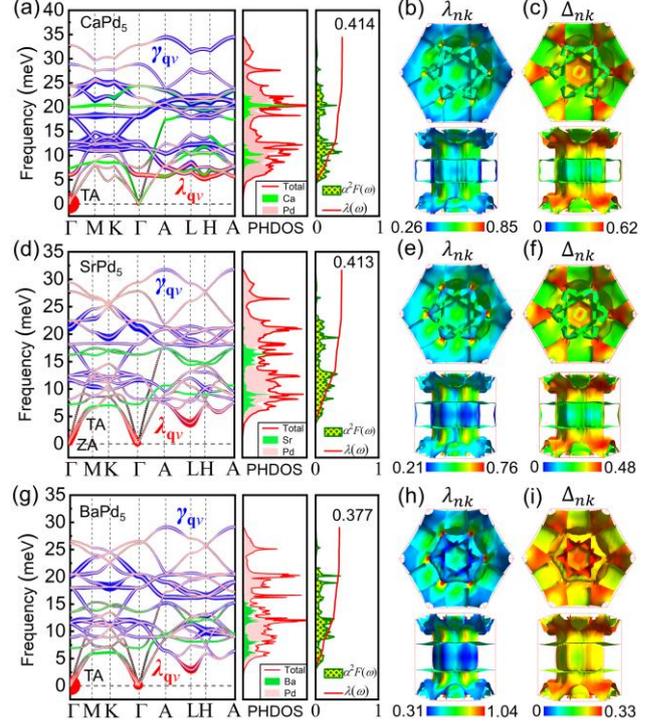

FIG. 2. Phonon dispersions with electron-phonon coupling $\lambda_{qv}$ (red balls) and phonon linewidth $\gamma_{qv}$ (blue balls), phonon dispersion weighted by the vibration modes of the atoms for M (green balls) and Pd (pink balls), projected phonon density of states (PHDOS), and frequency-dependent Eliashberg spectral functions $\alpha^2F(\omega)$ with cumulative EPC strength $\lambda(\omega)$. Top and side views of EPC $\lambda_{n\mathbf{k}}$ and the superconducting gap $\Delta_{n\mathbf{k}}$ on Fermi surfaces by using the FERMI-SURFER code [99]. (a)-(c) for CaPd$_5$, (d)-(f) for SrPd$_5$, (g)-(i) for BaPd$_5$.

Next, we only focus on the three stable CaPd$_5$, SrPd$_5$, and BaPd$_5$. As shown in Fig. 2, based on their element-projected phonon dispersion and phonon DOS (PHDOS), we can divide the vibration mode of MPd$_5$ into two parts: The range of 5~20 meV is contributed by a mixing of Pd and M vibrations, while the high-frequency range (> 20 meV) and low-frequency range (< 5 meV) are only dominated by Pd vibrations. The contribution of Pd vibrations to the phonon dispersion is larger in MPd$_5$, resulting in a pronounced phonon linewidth $\gamma_{qv}$ (at wave vector **q** and branch $v$). One can see that there are strong EPCs $\lambda_{qv}$ near the Γ point, and their largest $\lambda_{qv}$ appears in the transverse acoustic (TA) phonon mode along the Γ-M path. More detailed mode-resolved EPCs $\lambda_{qv}$ of CaPd$_5$ are shown in Fig. S10 [100]. The mode-resolved EPCs $\lambda_{qv}$ of SrPd$_5$ and BaPd$_5$ are shown in Figs. S11 and S12. The fundamental reason for the unique division of phonon dispersion in $M$Pd$_5$, lies in the fact that the $M$Pd$_5$ kagome structure is mainly composed of Pd$_5$, while the $M$ atoms make a predominant contribution to the stable structural framework.

To investigate their superconductivity, we calculated the EPC of $M$Pd$_5$. The right panel of Fig. 2(a) shows the calculated frequency-dependent Eliashberg electron-phonon spectrum function $\alpha^2F(\omega)$ and the cumulative EPC intensity $\lambda(\omega)$, where $\alpha^2F(\omega)$ exhibits sharp peaks due to the multimodal PHDOS and the dispersion of multiple flat band phonon dispersion. The steady rise of the $\alpha^2F(\omega)$ leads to a frequency-dependent cumulative EPC strength $\lambda(\omega)$ = 0.414. As shown in the right panels in Figs. 2(d) and 2(g), $\alpha^2F(\omega)$ and $\lambda(\omega)$ for SrPd$_5$ and BaPd$_5$ exhibit the similar upward trend as for CaPd$_5$. The steady rise of the curve at $\alpha^2F(\omega)$ results in a $\lambda(\omega)$ of 0.413 for SrPd$_5$ and 0.377 for BaPd$_5$. Note that $\lambda(\omega)$ is defined as:

$$\lambda(\omega) = \sum_{\mathbf{q}\nu} \lambda_{\mathbf{q}\nu} = 2\int_0^\omega \frac{\alpha^2 F(\omega)}{\omega} d\omega. \quad (1)$$

The $T_c$ value of the three materials at static zero pressure are obtained using the bisection method [91, 93, 94, 98], as shown in Table II. By comparing the three structures, we see that their EPC strength $\lambda(\omega)$ increases gradually as the M atoms change from Ba to Sr and Ca. The results show that the $T_c$ of the $M$Pd$_5$ materials follow the trend of $T_c$ (BaPd$_5$) < $T_c$ (SrPd$_5$) < $T_c$ (CaPd$_5$). We note that the $T_c$ of the $M$Pd$_5$ is lower than that of MgB$_3$ kagome material ($T_c$ = 12.2 K) [12], but higher than that of the (135)-type systems [24-30]. Therefore, $M$Pd$_5$ is expected to become a potential superconductor candidate.

$M$Pd$_5$ exhibits strong anisotropy and weak EPC superconductivity [see Figs. 2(b), 2(e), and 2(h)]. The $\lambda_{n\mathbf{k}}$ (band index $n$ and wave-number $\mathbf{k}$ dependent) of CaPd$_5$'s projection on the Fermi surface shows a ratio of max. to min. value of 3.3. The steep bands 48, 50 and 51 along with flat band 49 contribute significantly to the EPC. The EPCs of SrPd$_5$ and BaPd$_5$ also exhibit strong anisotropy, with $\lambda_{n\mathbf{k}}$ of 3.6 and 3.4, respectively. The projection of $\lambda_{n\mathbf{k}}$ of CaPd$_5$ on the four-band intersecting with $E_F$ can be found in Fig. S13 [100]. The mean values of EPC ($\lambda_{avg}$) for CaPd$_5$, SrPd$_5$, and BaPd$_5$ are 0.45, 0.40, and 0.52, respectively (< 1), indicating weak EPC superconductivity. The $\lambda_{avg}$ of the (15)-type $M$Pd$_5$ is slightly smaller than that of the (13)-type MgB$_3$ ($\lambda_{avg}$ = 0.63) [12] kagome material, while like MgB$_3$, all of them have the anisotropic EPC. The anisotropic EPC features of $M$Pd$_5$ lead to the anisotropic superconducting gap $\Delta_{n\mathbf{k}}$ [see Figs. 2(c), 2(f), and 2(i)]. For example, the $\Delta_{n\mathbf{k}}$ of CaPd$_5$, with an average value of 0.144 meV, displays a significant anisotropy. The projections of $\Delta_{n\mathbf{k}}$ on the four-band passing through $E_F$ is shown in Fig. S14 [100]. Note that the superconducting gap at a wave vector $\mathbf{k}$ is given by

$$\Delta_{n\mathbf{k}} = -\frac{1}{2}\sum_{n'\mathbf{k}'} \frac{K_{n\mathbf{k}n'\mathbf{k}'}(\varepsilon_{n\mathbf{k}},\varepsilon_{n'\mathbf{k}'})}{1+Z_{n\mathbf{k}}(\varepsilon_{n\mathbf{k}})} \times \frac{\Delta_{n'\mathbf{k}'}}{\sqrt{\varepsilon_{n'\mathbf{k}'}^2+\Delta_{n'\mathbf{k}'}^2}} \tanh(\frac{\sqrt{\varepsilon_{n'\mathbf{k}'}^2+\Delta_{n'\mathbf{k}'}^2}}{2T}), \quad (2)$$

## C. Quantum control of superconductivities

Generally, the following measures can be used to tune the superconductivtiy, including external pressure [105-107], doping [108], and the isotope effect [57]. Here, we examine the role of pressure and carrier doping in tuning $T_c$ of CaPd$_5$, the largest $T_c$ among the three materials under study. At both the max. and min. values of the pressure and carrier doping, the phonon spectra of CaPd$_5$ have no virtual frequency, indicating that they are dynamically stable (see Fig. S15 in Supplemental Material [100]). Applying an external hydrostatic pressure ranging from -10 to 10 GPa, we found that when positive pressure is applied, FB gradually moves upward with the increase of pressure, and the lattice constant (L), $N(E_F)$, $\lambda$, superconducting gap and $T_c$ all decrease monotonically [see Figs. 3(a), 3(c), and 3(e)]. Conversely, when negative pressure is applied, FB gradually moves downward with increasing pressure, and the lattice constants (L), $N(E_F)$, $\lambda$, superconducting gap, and $T_c$ all monotonically increase. This is because, for a conventional superconductor, the $T_c$ is closely related to the $N(E_F)$ (i.e., DOS at $E_F$ when in the superconducting state). The negative pressure may slightly increase the $T_c$ of CaPd$_5$ due to the increased $N(E_F)$, which is proportional to EPC $\lambda$. This implies the application of external pressure as a tuning knob for manipulating the superconductivity properties of $M$Pd$_5$.

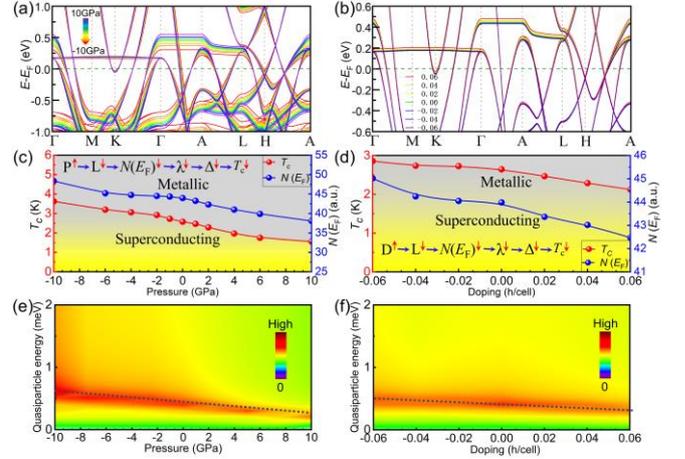

FIG. 3. Band diagram with (a) pressure variation and (b) carrier doping concentration of CaPd$_5$. $T_c$ and $N(E_F)$ with (c) pressure-dependent and (d) carrier doping concentration, as well as phase diagrams of superconducting and metallic states. Superconducting quasiparticle density of states (QPDOS) associated with (e) pressure-dependent and (f) carrier doping at 0.1 K. Black dashed lines depict trends of one-gap peak vs pressure or carrier doping.

To further evaluate the pressure-mediated behavior, we employ conductivity spectroscopy, an essential technique for probing the properties of superconductors, such as their surface states. Figure 3(c) depicts the pressure-dependent superconducting quasiparticle density of states (QPDOS) of CaPd$_5$ at 0.1 K as a function of applied pressure, which is directly relevant for comparison with experimental tunneling conductance. The presence of a single QPDOS peak confirms that $CaPd_5$ behaves as a one-gap superconductor, where the peak at the quasiparticle energy corresponds to its superconducting gap. Under positive pressure, the amplitude of QPDOS peak decreases and shifts toward low energy. Conversely, applying negative pressure results in a broadened and intensified QPDOS peak with the increasing pressure, shifting towards higher energy. These findings indicate that $T_c$ of $CaPd_5$ can be modulated and effectively enhanced through the application of negative pressure.

Furthermore, to enhance $T_c$ of $CaPd_5$, we also explore the effect of carrier doping on $N(E_F)$ and the corresponding shift of $E_F$. We employ a jellium model that simulates hole and electron doping through direct modulation of the total number of electrons in the system. The corresponding band structure under various carrier doping concentrations is depicted in Fig. 3(b). The increase of electron doping progressively shifts the FB downward, whereas increase in hole doping concentration shifts it upward. As expected, $N(E_F)$ gradually increases with higher electron concentration, while it conversely decreases monotonically with increasing hole doping concentration. As a result, the $\lambda$, superconducting gap, and $T_c$ exhibit the same behavior as that observed in $N(E_F)$ [see Fig. 3(d)]. For instance, $T_c$ can be increased to 2.85 K at a high electron doping concentration of 0.06 $e$ per cell. In Fig. 3(f), we present the superconducting QPDOS associated with carrier doping at 0.1 K. As the hole doping concentration rises, the QPDOS peak moves toward lower energy, whereas increased electron doping shift it toward higher energy. These findings confirm that electron doping is an effective mechanism for enhancing the $T_c$ of $CaPd_5$, while hole doping diminishes it.

### D. Nontrivial topological properties

TABLE III. Topological and node superconductor $CaPd_5$ diagnosis based on symmetry. The first column is the point group of $CaPd_5$. The second column indicates the cases I-IV for each pairing. Symbols [L] and [P] in case I specify the shape of line (L) and point (P), respectively. The fourth column represents the paths where CRs are broken for case I and the entry of symmetry indicators for case II. For case II of $CaPd_5$, the entries are in $(\mathbb{Z}_2)^4 \times \mathbb{Z}_{12} \times \mathbb{Z}_{24}$ [109].

| PG | Pairing | Case | Nodes | Topology |
|---|---|---|---|---|
| $D_{6h}$ | $B_{1u}$ | I [P] | $\Gamma$-M, A-L, $\Gamma$-A | NSC |
| | $B_{1g}$ | I [L] | $\Gamma$-K, $\Gamma$-M, A-L, L-H, $\Gamma$-A, M-L | NSC |
| | $B_{2u}$ | I [P] | $\Gamma$-K, L-H, $\Gamma$-A | NSC |
| | $B_{2g}$ | I [L] | $\Gamma$-K, $\Gamma$-M, A-L, L-H, $\Gamma$-A, M-L | NSC |
| | $A_{2u}$ | I [P] | $\Gamma$-A, M-L | NSC |
| | $A_{2g}$ | I [L] | $\Gamma$-K, $\Gamma$-L, A-L, L-H, $\Gamma$-A, M-L | NSC |
| | $A_{1u}$ | II | (0, 1, 1, 1, 7, 13) | TSC or TNSC |
| | $A_{1g}$ | IV | … | … |

We now turn to the topological properties of the $MPd_5$ systems. Notably, $CaPd_5$ exhibits a non-trivial topological property with the index of $Z_4 = 1$, defined by the sum of the inverted parities of all occupied bands at all eight points with the time-reversal invariant momenta. In contrast, the other two members of the $MPd_5$ family display trivial topological indices, implying their non-topological nature. This conclusion was reached using the QEIRREPS code [86] in combination with the SOC results from the QE code [74]. We further investigate the superconductor property of $CaPd_5$ using symmetry indicators (SIs) [83-85] to determine whether it behaves as a topological superconductor, a nodal superconductor, or simply a conventional superconductor. This method, mainly employed to examine the irreducible representations of the space groups, has been extended to various materials, including superconductors [109-112]. Table III presents a quantum-mechanics based diagnosis of the topological and node-superconductor properties of $CaPd_5$, including the SOC effect, obtained using the QEIRREPS [86] and TOPOLOGICAL SUPERCON tools [109, 113]. For each pairing symmetry, superconductors fall into the following categories: case I, representation-enforced nodal superconductor (NSC); case II, topological NSC (TNSC) or symmetry-diagnosable topological superconductor (TSC); case III, topology-trivial or not symmetry-diagnosable TSC; and case IV, silent for the trivial pairing (due to the fact that no band labels can be defined at any high-symmetry momenta).

Their point group $D_{6h}$ contains eight 1D single-valued representations: $B_{1u}$, $B_{1g}$, $B_{2u}$, $B_{2g}$, $A_{2u}$, $A_{2g}$, $A_{2u}$, and $A_{2g}$, all of which preserve time-reversal symmetry (see Table III). Our analysis predicts that the superconducting phase of $CaPd_5$ has nodal lines (L) in the case of *even*-parity and nodal points (P) in the case of *odd*-parity. We find that a superconductor with $B_{1u}$, $B_{1g}$, $B_{2u}$, $B_{2g}$, $A_{2u}$, or $A_{2g}$ pairing is a representation-enforced NSC. The compatibility relations (CRs, or symmetry constrains) along various lines are violated, as shown in Table III. For example, in the case of $B_{1u}$ pairing, the superconductor is an NSC with point-like nodes, where the CRs are broken along the $\Gamma$-M, A-L, $\Gamma$-A paths. In the case of $A_{1u}$ pairing, all CRs are satisfied. Therefore, the SIs method [83-85] is suitable to diagnose topology. We find that the material belongs to the entry (0, 1, 1, 1, 7, 13) $\in (\mathbb{Z}_2)^4 \times \mathbb{Z}_{12} \times \mathbb{Z}_{24}$ of SIs, where the first four components (0, 1, 1, 1) $\in (\mathbb{Z}_2)^4$ represent the topological invariants under time-reversal symmetry and other fundamental symmetry operations, while the fifth component (7) $\in \mathbb{Z}_{12}$ and the sixth component (13) $\in \mathbb{Z}_{24}$ correspond to distinct topological invariants associated with specific periodic symmetries. These components can be transformed into mirror Chern numbers through modular operations, which serve as crucial topological invariants for characterizing the topological properties of materials under mirror symmetry. Finally, the trivial $A_{1g}$ pairing symmetry is categorized under case IV, where no well-defined band labels exist [109]. Notably, none of the eight pairing symmetries of CaPd$_5$ exhibit case III which represents trivial or undiagnosable superconducting phases.

We proceed with further analysis of the topological properties of CaPd$_5$ material using the WANNIER 90 [87] and WANNLER TOOLS [88] programs. The results reveal a rich array of topologically non-trivial Dirac crossing points (DC) near the $E_F$. Among these, DC$_1$, DC$_2$, and DC$_3$ form three loops in the $k_x$-$k_y$ plane, as illustrated in Figs. 4(a), 4(b), and 4(c). Specifically, along the high-symmetric line Γ-$M$ in the Brillouin zone (BZ), DC$_1$ is formed by the states SM$_4$ and SM$_2$, while DC$_2$ is formed by the states SM$_4$ and SM$_1$, as shown in Fig. 4(e). Due to the constraints of electron movement imposed by crystal symmetry, we represent these states through the irreducible representations of symmetric operation groups derived from irvsp [89]. For the DC$_1$ and DC$_2$ at Γ-$M$, they belong to the SM small group, which is characterized by four symmetry operations and four irreducible representations, as summarized in Table IV. The states forming the DC$_1$ point can be represented by SM$_4$ and SM$_2$, respectively. On the left side of DC$_1$ (Γ-DC$_1$), energy of state SM$_2$ is higher than SM$_4$; however, to the right of DC$_1$ (DC$_1$-$M$), this relationship is reversed. A similar pattern is observed for DC$_2$, which also consists of two states, represented by SM$_4$ and SM$_2$, with an inverted energy order across the Γ-DC$_2$ and DC$_2$-$M$ segments. Along the Γ-$K$ line, DC$_1$ and DC$_2$ points belong to a different small group, designated as LD, whose symmetry operations and irreducible representations are detailed in Table IV. As illustrated in Fig. 4(f), both DC$_1$ and DC$_2$ are formed of LD$_4$ and LD$_2$. The energy order reversed also happened at DC$_1$ and DC$_2$.

TABLE IV. The symmetric operation (Oper.) matrix and irreducible representation of SM and LD small groups obtained by the Bilbao Crystallographic Server [114]. The first column is crystal symmetry operations. Columns 2 through 9 are one-dimensional irreducible representations.

| Oper. | SM$_1$ | SM$_2$ | SM$_3$ | SM$_4$ | LD$_1$ | LD$_2$ | LD$_3$ | LD$_4$ |
|---|---|---|---|---|---|---|---|---|
| $E$ | 1 | 1 | 1 | 1 | 1 | 1 | 1 | 1 |
| $2_{210}$ | 1 | -1 | -1 | 1 | 1 | 1 | -1 | -1 |
| $m_{001}$ | 1 | 1 | -1 | -1 | 1 | -1 | -1 | 1 |
| $m_{010}$ | 1 | -1 | 1 | -1 | | | | |
| $m_{1\text{-}10}$ | | | | | 1 | -1 | 1 | -1 |

In the $k_x$-$k_y$ plane, the constrain symmetry operations are reduced to only two: $E$ and $m_{001}$. Under this condition, we can only use the eigenvalue of $m_{001}$ to represent the energy order. For loops formed by DC$_1$, DC$_2$, or DC$_3$, the energy order within the inner space of a loop is opposite to that of the outer. These loops consist of energy-reversed points, and their topological nontrivial structure is a characteristic feature. Topological surface states provide evidence for the non-trivial nodal structure inherent in materials. We compute the surface states [see Fig. 4(d)] using Green's functionmethod with a tight-binding Hamiltonian in the WANNIER 90 [87] package. For the (1 -1 0) surfaces with the energy range of 5 eV near the $E_F$, the surface states along the connection Γ$_0$-Γ$_1$ are found to emerge around -0.5 eV. These findings demonstrate the nontrivial topological properties of CaPd$_5$. We note that the topologically protected exotic surface states demonstrated here can be harnessed for the development of advanced topological superconducting devices and quantum circuits, which have significant implications for quantum information technology.

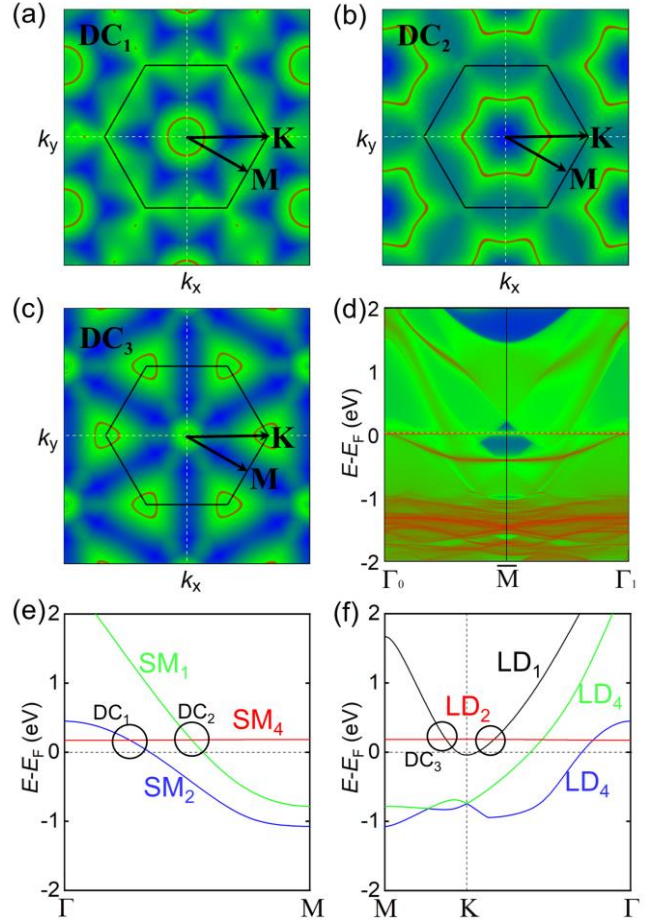

FIG. 4. (a)-(c) The three loops formed by DC$_1$, DC$_2$ and DC$_3$ respectively, and (d) depicts the surface state at $E_F$ on the semi-infinite (1-10) surface. (e) and (f) illustrate Dirac points DC$_1$, DC$_2$, and DC$_3$ formed by different states along the high-symmetric line of BZ.

## IV. CONCLUSIONS

In summary, we have predicted and explored the $M$Pd$_5$ kagome materials, focusing on their topological properties, and superconductivities through first-principles calculations. The results of phonon spectra, elastic constants, convex-hull energy, and molecular dynamics simulations confirm their dynamic, mechanical, and thermodynamic stability. Based on the density functional thery for superconductors, we predict that $M$Pd$_5$ ($M$ = Ca, Sr, Ba) materials are anisotropic one-gap superconductors with weak EPC. The calculated results show that the EPC strength $\lambda(\omega)$ of CaPd$_5$/SrPd$_5$/BaPd$_5$ is 0.414/0.413/0.377, and the corresponding $T_c$ is 2.64, 2.03, and 1.50 K, respectively. We investigate the effects of external pressure and carrier doping on the superconductivity of CaPd$_5$ and find that $T_c$ can be effectively enhanced by applying negative pressure and electron doping. For example, CaPd$_5$ possesses non-trivial topological properties based on its $Z_4 = 1$, and rich topological non-trivial crossing points near the $E_F$ and the surface states of (1-10) surfaces clearly indicate that the kagome material is topological metal. These insights position this kagome $M$Pd$_5$ family of materials as a compelling platform for exploring exotic kagome physics, advancing the search for superconductors, topological materials, and functional materials with (15)-type kagome structures.

## ACKNOWLEDGMENTS

This work was supported by the Natural Science Foundation of Henan Province (Grant No. 242300421214), the National Natural Science Foundation of China (Grant s No. 12274117, No. 61835013, No. 12174461, No. 62274066, No. 62275074, No. 12334012, No. 52327808, and No. 12234012), the National Key R&D Program of China (Grants No. 2021YFA1400900, No. 2021YFA0718300, No. 2024YFF0726700, and No. 2021YFA1402100), the Program for Innovative Research Team (in Science and Technology) in University of Henan Province (Grant No. 24IRTSTHN025), and Space Application System of China Manned Space Program, Henan Center for Outstanding Overseas Scientists (No. GZS2023007). We thank the High-Performance Computing Center of Henan Normal University.

## DATA AVAILABILITY

The data that support the findings of this article are not publicly available. The data are available from the authors upon reasonable request.

[1] I. Syôzi, Statistics of kagomé lattice, Prog. Theor. Phys. **6**, 306 (1951).

[2] T. Neupert, M. M. Denner, J.-X. Yin, R. Thomale, and M. Z. Hasan, Charge order and superconductivity in kagome materials, Nat. Phys. **18**, 137 (2022).

[3] J.-X. Yin, B. Lian, and M. Z. Hasan, Topological kagome magnets and superconductors, Nature **612**, 647 (2022).

[4] S. Kumar Pradhan, S. Pradhan, P. Mal, P. Rambabu, A. Lakhani, B. Das, B. Lingam Chittari, G. R. Turpu, and P. Das, Endless dirac nodal lines and high mobility in kagome semimetal Ni$_3$In$_2$Se$_2$: a theoretical and experimental study, J. Phys.:Condens. Matter **36**, 445601 (2024).

[5] H. Ma, G. H. Liu, L. Yang, Q. W. Yin, P. F. Ding, Z. H. Liu, H. Y. Lu, H. C. Lei, and S. C. Wang, Angle-resolved photoemission study of V-based kagome metal CsV$_8$Sb$_{12}$, Phys. Rev. B **108**, 235129 (2023).

[6] Y.-X. Jiang, J.-X. Yin, M. M. Denner, N. Shumiya, B. R. Ortiz, G. Xu, Z. Guguchia, J. He, M. S. Hossain, X. Liu, J. Ruff, L. Kautzsch, S. S. Zhang, G. Chang, I. Belopolski, Q. Zhang, T. A. Cochran, D. Multer, M. Litskevich, Z.-J. Cheng *et al*., Unconventional chiral charge order in kagome superconductor KV$_3$Sb$_5$, Nat. Mater. **20**, 1353 (2021).

[7] L. Ye, M. Kang, J. Liu, F. von Cube, C. R. Wicker, T. Suzuki, C. Jozwiak, A. Bostwick, E. Rotenberg, D. C. Bell, L. Fu, R. Comin and J. G. Checkelsky, Massive dirac fermions in a ferromagnetic kagome metal, Nature **555**, 638 (2018).

[8] W. R. Meier, J. Yan, M. A. McGuire, X. P. Wang, A. D. Christianson, and B. C. Sales, Reorientation of antiferromagnetism in cobalt doped FeSn, Phys. Rev. B **100**, 184421 (2019).

[9] W. R. Meier, M. H. Du, S. Okamoto, N. Mohanta, A. F. May, M. A. McGuire, C. A. Bridges, G. D. Samolyuk, and B. C. Sales, Flat bands in the CoSn-type compounds, Phys. Rev. B **102**, 075148 (2020).

[10] B. C. Sales, W. R. Meier, A. F. May, J. Xing, J. Q. Yan, S. Gao, Y. H. Liu, M. B. Stone, A. D. Christianson, Q. Zhang and M. A. McGuire, Tuning the flat bands of the kagome metal CoSn with Fe, In, or Ni doping, Phys. Rev. Mater. **5**, 044202 (2021).

[11] H. Huang, L. X. Zheng, Z. Y. Lin, X. Guo, S. Wang, S. Zhang, C. Zhang, Z. Sun, Z. F. Wang, H. M. Weng, L. Li, T. Wu, X. H. Chen and C. G. Zeng, Flat-band-induced anomalous anisotropic charge transport and orbital magnetism in kagome metal CoSn, Phys. Rev. Lett. **128**, 096601 (2022).

[12] Y. P. An, J. C. Chen, Z. X. Wang, J. Li, S. J. Gong, C. L. Ma, T. X. Wang, Z. Y. Jiao, R. Q. Wu, J. P. Hu and W. M. Liu, Topological and nodal superconductor kagome magnesium triboride, Phys. Rev. Mater. **7**, 014205 (2023).

[13] L. Yang, Y. P. Li, H. D. Liu, N. Jiao, M. Y. Ni, H. Y. Lu, P. Zhang, and C. S. Ting, Theoretical prediction of superconductivity in boron kagome monolayer: MB$_3$ (M = Be, Ca, Sr) and the hydrogenated CaB$_3$, Chin. Phys. Lett. **40**, 017402 (2023).

[14] Z. Y. Qu, F. J. J. Han, T. Yu, M. L. Xu, Y. W. Li, and G. C. Yang, Boron kagome-layer induced intrinsic superconductivity in a MnB$_3$ monolayer with a high critical temperature, Phys. Rev. B **102**, 075431 (2020).

[15] S. Y. Xie, X. B. Li, W. Q. Tian, N. K. Chen, Y. L. Wang, S. B. Zhang, and H. B. Sun, A novel two-dimensional MgB$_6$ crystal: metal-layer stabilized boron kagome lattice, Phys. Chem. Chem. Phys. **17**, 1093 (2015).

[16] T. Bo, P. F. Liu, L. Yan, and B. T. Wang, Electron-phonon coupling superconductivity in two-dimensional orthorhombic MB$_6$ (M = Mg, Ca, Ti, Y) and hexagonal MB$_6$ (M = Mg, Ca, Sc, Ti), Phys. Rev. Mater. **4**, 114802 (2020).

[17] G. Profeta, M. Calandra, and F. Mauri, Phonon-mediated superconductivity in graphene by lithium deposition, Nat. Phys. **8**, 131 (2012).

[18] A. Zelenskiy, T. L. Monchesky, M. L. Plumer, and B. W. Southern, Anisotropic magnetic interactions in hexagonal AB-stacked kagome lattice structures: Application to $Mn_3X$ (X = Ge,Sn,Ga) compounds, Phys. Rev. B **103**, 144401 (2021).

[19] D. Y. Zhang, Z. P. Hou, and W. B. Mi, Progress in magnetic alloys with kagome structure: materials, fabrications and physical properties, J. Mater. Chem. C **10**, 7748 (2022).

[20] B. Mortazavi, X. Y. Zhuang, and T. Rabczuk, A first-principles study on the physical properties of two-dimensional $Nb_3Cl_8$, $Nb_3Br_8$ and $Nb_3I_8$, Appl. Phys. A:Mater. Sci. Process. **128**, 934 (2022).

[21] Y. L. Feng, and Q. Yang, Enabling triferroics coupling in breathing kagome lattice $Nb_3X_8$ (X = Cl, Br, I) monolayers, J. Mater. Chem. C **11**, 5762 (2023).

[22] C. Mielke, Y. Qin, J. X. Yin, H. Nakamura, D. Das, K. Guo, R. Khasanov, J. Chang, Z. Q. Wang, S. Jia, S. Nakatsuji, A. Amato, H. Luetkens, G. Xu, M. Z. Hasan and Z. Guguchia, Nodeless kagome superconductivity in $LaRu_3Si_2$, Phys. Rev. Mater. **5**, 034803 (2021).

[23] X. Gui, and R. J. Cava, $LaIr_3Ga_2$: a superconductor based on a kagome lattice of Ir, Chem. Mater. **34**, 2824 (2022).

[24] B. R. Ortiz, L. C. Gomes, J. R. Morey, M. Winiarski, M. Bordelon, J. S. Mangum, I. W. H. Oswald, J. A. Rodriguez-Rivera, J. R. Neilson, S. D. Wilson, E. Ertekin, T. M. McQueen and E. S. Toberer, New kagome prototype materials: discovery of $KV_3Sb_5$, $RbV_3Sb_5$, and $CsV_3Sb_5$, Phys. Rev. Mater. **3**, 094407 (2019).

[25] Y. H. Gu, Y. Zhang, X. L. Feng, K. Jiang, and J. P. Hu, Gapless excitations inside the fully gapped kagome superconductors $AV_3Sb_5$, Phys. Rev. B **105**, L100502 (2022).

[26] J. G. Si, W. J. Lu, Y. P. Sun, P. F. Liu, and B. T. Wang, Charge density wave and pressure-dependent superconductivity in the kagome metal $CsV_3Sb_5$: a first-principles study, Phys. Rev. B **105**, 024517 (2022).

[27] S. W. Kim, H. Oh, E. Moon, and Y. Kim, Monolayer kagome metals $AV_3Sb_5$, Nat. Commun. **14**, 591 (2023).

[28] S. D. Wilson, and B. R. Ortiz, $AV_3Sb_5$ kagome superconductors, Nat. Rev. Mater. **9**, 420 (2024).

[29] B. R. Ortiz, P. M. Sarte, E. M. Kenney, M. J. Graf, S. M. L. Teicher, R. Seshadri, and S. D. Wilson, Superconductivity in the $Z_2$ kagome metal $KV_3Sb_5$, Phys. Rev. Mater. **5**, 034801 (2021).

[30] B. R. Ortiz, S. M. L. Teicher, Y. Hu, J. L. Zuo, P. M. Sarte, E. C. Schueller, A. M. M. Abeykoon, M. J. Krogstad, S. Rosenkranz, R. Osborn, R. Seshadri, L. Balents, J. He and S. D. Wilson, $CsV_3Sb_5$: a $Z_2$ topological kagome metal with a superconducting ground state, Phys. Rev. Lett. **125**, 247002 (2020).

[31] L. T. Shi, J. G. Si, A. Liang, R. Turnbull, P. F. Liu, and B. T. Wang, Topological and superconducting properties in bilayer kagome metals $YT_6Sn_6$ (T = V, Nb, Ta), Phys. Rev. B **107**, 184503 (2023).

[32] M. Z. Shi, F. H. Yu, Y. Yang, F. B. Meng, B. Lei, Y. Luo, Z. Sun, J. F. He, R. Wang, Z. C. Jiang, Z. T. Liu, D. W. Shen, T. Wu, Z. Y. Wang, Z. J. Xiang, J. J. Ying and X. H. Chen, A new class of bilayer kagome lattice compounds with dirac nodal lines and pressure-induced superconductivity, Nat. Commun. **13**, 2773 (2022).

[33] Y. Yang, R. Wang, M. Z. Shi, Z. Wang, Z. Xiang, and X. H. Chen, Type-II nodal line fermions in the $Z_2$ topological semimetals $AV_6Sb_6$ (A = K, Rb, and Cs) with a kagome bilayer, Phys. Rev. B **104**, 245128 (2021).

[34] D. F. Liu, A. J. Liang, E. K. Liu, Q. N. Xu, Y. W. Li, C. Chen, D. Pei, W. J. Shi, S. K. Mo, P. Dudin, T. Kim, C. Cacho, G. Li, Y. Sun, L. X. Yang, Z. K. Liu, S. S. P. Parkin, C. Felser and Y. L. Chen, Magnetic weyl semimetal phase in a kagomé crystal, Science **365**, 1282 (2019).

[35] G. Sala, J. Y. Y. Lin, A. M. Samarakoon, D. S. Parker, A. F. May, and M. B. Stone, Ferrimagnetic spin waves in honeycomb and triangular layers of $Mn_3Si_2Te_6$, Phys. Rev. B **105**, 214405 (2022).

[36] D. Huang, H. Nakamura, K. Küster, U. Wedig, N. B. M. Schröter, V. N. Strocov, U. Starke, and H. Takagi, Probing the interlayer coupling in $2H$-$NbS_2$ via soft x-ray angle-resolved photoemission spectroscopy, Phys. Rev. B **105**, 245145 (2022).

[37] C. Mielke Iii, H. Liu, D. Das, J. X. Yin, L. Z. Deng, J. Spring, R. Gupta, M. Medarde, C. W. Chu, R. Khasanov, Z. M. Hasan, Y. Shi, H. Luetkens and Z. Guguchia, Local spectroscopic evidence for a nodeless magnetic kagome superconductor $CeRu_2$, J. Phys.:Condens. Matter **34**, 485601 (2022).

[38] M. Pereiro, D. Yudin, J. Chico, C. Etz, O. Eriksson, and A. Bergman, Topological excitations in a kagome magnet, Nat. Commun. **5**, 4815 (2014).

[39] S. Shao, J. X. Yin, I. Belopolski, J. Y. You, T. Hou, H. Y. Chen, Y. X. Jiang, M. S. Hossain, M. Yahyavi, C. H. Hsu, Y. P. Feng, A. Bansil, M. Z. Hasan and G. Q. Chang, Intertwining of magnetism and charge ordering in kagome FeGe, ACS Nano **17**, 10164 (2023).

[40] X. K. Teng, J. S. Oh, H. X. Tan, L. B. Chen, J. W. Huang, B. Gao, J. X. Yin, J. H. Chu, M. Hashimoto, D. H. Lu, C. Jozwiak, A. Bostwick, E. Rotenberg, G. E. Granroth, B. Yan, R. J. Birgeneau, P. C. Dai and M. Yi, Magnetism and charge density wave order in kagome FeGe, Nat. Phys. **19**, 814 (2023).

[41] M. Wenzel, E. Uykur, A. A. Tsirlin, S. Pal, R. M. Roy, C. Yi, C. Shekhar, C. Felser, A. V. Pronin, and M. Dressel, Intriguing low-temperature phase in the antiferromagnetic kagome metal FeGe, Phys. Rev. Lett. **132**, 266505 (2024).

[42] S. F. Wu, M. L. Klemm, J. Shah, E. T. Ritz, C. R. Duan, X. K. Teng, B. Gao, F. Ye, M. Matsuda, F. K. Li, X. H. Xu, M. Yi, T. Birol, P. C. Dai and G. Blumberg, Symmetry breaking and ascending in the magnetic kagome metal FeGe, Phys. Rev. X **14**, 011043 (2024).

[43] B. H. Zhang, J. Y. Ji, C. S. Xu, and H. J. Xiang, Electronic and magnetic origins of unconventional charge density wave in kagome FeGe, Phys. Rev. B **110**, 125139 (2024).

[44] S. S. Philip, J. Yang, D. Louca, P. F. S. Rosa, J. D. Thompson, and K. L. Page, Bismuth kagome sublattice distortions by quenching and flux pinning in superconducting $RbBi_2$, Phys. Rev. B **104**, 104503 (2021).

[45] W. Qin, B. Xu, S. Liao, G. Liu, B. Sun, and M. Wu, Flat-band splitting induced tunable magnetism in defective $CrI_3$ monolayer, Solid State Commun. **321**, 114037 (2020).

[46] X. R. Li, Y. Dai, Y. D. Ma, Q. L. Sun, W. Wei, and B. B. Huang, Exotic quantum spin hall effect and anisotropic spin splitting in carbon based $TMC_6$ (TM = Mo, W) kagome monolayers, Carbon **109**, 788 (2016).

[47] H. J. Kim, M. J. Kim, J. Lee, J. M. Ok, and C.-J. Kang, Tuning the flat band with in-plane biaxial strain and the emergence of superconductivity in $Ni_3Sn$, Phys. Rev. B **110**, 024504 (2024).

[48] S. Xiao, D. Zhang, and N. Wang, Giant anomalous hall and nernst conductivities in cubic $Fe_3Pt$ and $Ni_3In$, J. Phys. D: Appl. Phys. **56**, 454001 (2023).

[49] S. Baidya, A. V. Mallik, S. Bhattacharjee, and T. Saha-Dasgupta, Interplay of magnetism and topological superconductivity in bilayer kagome metals, Phys. Rev. Lett. **125**, 026401 (2020).

[50] J. T. Yang, C. Jing, Y. C. Xiong, and S. J. Luo, Electronic structures and magnetic properties of two-dimensional honeycomb-kagome structured $V_2O_3$ monolayer and $V_2O_3/M(0001)$ (M = Zr and Hf) system, J. Magn. Magn. Mater. **564**, 170161 (2022).

[51] Y. C. Zhao, M. X. Zhu, Y. Wang, and P. Li, Honeycomb-kagome lattice $Na_3Te_2$: Dirac half-metal with quantum anomalous hall effect, Chem. Phys. **562**, 111658 (2022).

[52] J. G. Si, L. T. Shi, B. Z. Chen, H. H. Yang, J. Y. Xu, M. Liu, and S. Meng, Coexistence of superconductivity and topological phase in kagome metals $ANb_3Bi_5$ (A = K, Rb, Cs), npj Comput. Mater. **10**, 96 (2024).

[53] S. T. Peng, Y. L. Han, G. Pokharel, J. C. Shen, Z. Y. Li, M. Hashimoto, D. H. Lu, B. R. Ortiz, Y. Luo, H. C. Li, M. Y. Guo, B. Q. Wang, S. T. Cui, Z. Sun, Z. H. Qiao, S. D. Wilson and J. F. He, Realizing kagome band structure in two-dimensional kagome surface states of $RV_6Sn_6$ (R = Gd, Ho), Phys. Rev. Lett. **127**, 266401 (2021).

[54] J. X. Li, J. X. Liu, S. A. Baronett, M. F. Liu, L. Wang, R. H. Li, Y. Chen, D. Z. Li, Q. Zhu, and X. Q. Chen, Computation and data driven discovery of topological phononic materials, Nat. Commun. **12**, 1204 (2021).

[55] P. M. Neves, J. P. Wakefield, S. Fang, H. Nguyen, L. Ye, and J. G. Checkelsky, Crystal net catalog of model flat band materials, npj Comput. Mater. **10**, 39 (2024).

[56] N. Regnault, Y. F. Xu, M. R. Li, D. S. Ma, M. Jovanovic, A. Yazdani, S. S. P. Parkin, C. Felser, L. M. Schoop, N. P. Ong, R. J. Cava, L. Elcoro, Z. D. Song and B. A. Bernevig, Catalogue of flat-band stoichiometric materials, Nature **603**, 824 (2022).

[57] Y. P. An, J. C. Chen, Y. Yan, J. F. Wang, Y. N. Zhou, Z. X. Wang, C. L. Ma, T. X. Wang, R. Q. Wu, and W. M. Liu, Higher-order topological and nodal superconducting transition-metal sulfides MS (M = Nb and Ta), Phys. Rev. B **108**, 054519 (2023).

[58] T. Y. Yang, Q. Wan, J. P. Song, Z. Du, J. Tang, Z. W. Wang, N. C. Plumb, M. Radovic, G. W. Wang, G. Y. Wang, Z. Sun, J. X. Yin, Z. H. Chen, Y. B. Huang, R. Yu, M. Shi, Y. M. Xiong and N. Xu, Fermi-level flat band in a kagome magnet, Quantum Front. **1**, 14 (2022).

[59] H. D. Scammell, J. Ingham, T. Li, and O. P. Sushkov, Chiral excitonic order from twofold van Hove singularities in kagome metals, Nat. Commun. **14**, 605 (2023).

[60] Y. Hu, X. X. Wu, B. R. Ortiz, S. L. Ju, X. Han, J. Z. Ma, N. C. Plumb, M. Radovic, R. Thomale, S. D. Wilson, A. P. Schnyder and M. Shi, Rich nature of van hove singularities in kagome superconductor $CsV_3Sb_5$, Nat. Commun. **13**, 2220 (2022).

[61] Y. Hu, X. X. Wu, Y. Q. Yang, S. Y. Gao, N. C. Plumb, A. P. Schnyder, W. W. Xie, J. Z. Ma, and M. Shi, Tunable topological Dirac surface states and van Hove singularities in kagome metal $GdV_6Sn_6$, Sci. Adv. **8**, eadd2024 (2022).

[62] A. Low, T. K. Bhowmik, S. Ghosh, and S. Thirupathaiah, Anisotropic nonsaturating magnetoresistance observed in $HoMn_6Ge_6$: A kagome Dirac semimetal, Phys. Rev. B **109**, 195104 (2024).

[63] A. Simon, Superconductivity and chemistry, Angew. Chem. Int. Ed. Engl. **36**, 1788 (1997).

[64] S. Q. Deng, A. Simon, and J. Köhler, Flat band–steep band scenario and superconductivity – the case of calcium, Solid State Sci. **2**, 31 (2000).

[65] D. Zhou, Q. Li, Y. M. Ma, Q. L. Cui, and C. F. Chen, Pressure-induced superconductivity in SnTe: A first-principles study, J. Phys. Chem. C **117**, 12266 (2013).

[66] S. M. Yan, D. A. Huse, and S. R. White, Spin-liquid ground state of the S = 1/2 kagome Heisenberg antiferromagnet, Science **332**, 1173 (2011).

[67] T. H. Han, J. S. Helton, S. Y. Chu, D. G. Nocera, J. A. Rodriguez-Rivera, C. Broholm, and Y. S. Lee, Fractionalized excitations in the spin-liquid state of a kagome-lattice antiferromagnet, Nature **492**, 406 (2012).

[68] H. L. Luo, Q. Gao, H. X. Liu, Y. H. Gu, D. S. Wu, C. J. Yi, J. J. Jia, S. L. Wu, X. Y. Luo, Y. Xu, L. Zhao, Q. Y. Wang, H. Q. Mao, G. D. Liu, Z. H. Zhu, Y. G. Shi, K. Jiang, J. P. Hu, Z. Y. Xu and X. J. Zhou, Electronic nature of charge density wave and electron-phonon coupling in kagome superconductor $KV_3Sb_5$, Nat. Commun. **13**, 273 (2022).

[69] L. P. Nie, K. L. Sun, W. R. Ma, D. W. Song, L. X. Zheng, Z. W. Liang, P. Wu, F. H. Yu, J. Li, M. Shan, D. Zhao, S. J. Li, B. L. Kang, Z. M. Wu, Y. B. Zhou, K. Liu, Z. J. Xiang, J. J. Ying, Z. Y. Wang, T. Wu and X. H. Chen, Charge-density-wave-driven electronic

nematicity in a kagome superconductor, Nature **604**, 59 (2022).

[70] R. Q. Ku, L. Yan, J. G. Si, S. Y. Zhu, B. T. Wang, Y. D. Wei, K. J. Pang, W. Q. Li, and L. J. Zhou, Ab initio investigation of charge density wave and superconductivity in two-dimensional Janus 2H/1T-MoSH monolayers, Phys. Rev. B **107**, 064508 (2023).

[71] M. Tuniz, A. Consiglio, D. Puntel, C. Bigi, S. Enzner, G. Pokharel, P. Orgiani, W. Bronsch, F. Parmigiani, V. Polewczyk, P. D. C. King, J. W. Wells, I. Zeljkovic, P. Carrara, G. Rossi, J. Fujii, I. Vobornik, S. D. Wilson, R. Thomale, T. Wehling *et al.*, Dynamics and resilience of the unconventional charge density wave in $ScV_6Sn_6$ bilayer kagome metal, Commun. Mater. **4**, 103 (2023).

[72] X. K. Li, J. Koo, Z. W. Zhu, K. Behnia, and B. H. Yan, Field-linear anomalous Hall effect and Berry curvature induced by spin chirality in the kagome antiferromagnet $Mn_3Sn$, Nat. Commun. **14**, 1642 (2023).

[73] D. Golovanova, H. X. Tan, T. Holder, and B. H. Yan, Tuning the intrinsic spin Hall effect by charge density wave order in topological kagome metals, Phys. Rev. B **108**, 205203 (2023).

[74] P. Giannozzi, O. Andreussi, T. Brumme, O. Bunau, M. Buongiorno Nardelli, M. Calandra, R. Car, C. Cavazzoni, D. Ceresoli, M. Cococcioni, N. Colonna, I. Carnimeo, A. Dal Corso, S. de Gironcoli, P. Delugas, R. A. DiStasio, A. Ferretti, A. Floris, G. Fratesi, G. Fugallo *et al.*, Advanced capabilities for materials modelling with Quantum ESPRESSO, J. Phys.:Condens. Matter **29**, 465901 (2017).

[75] D. R. Hamann, Optimized norm-conserving Vanderbilt pseudopotentials, Phys. Rev. B **88**, 085117 (2013).

[76] M. Schlipf, and F. Gygi, Optimization algorithm for the generation of ONCV pseudopotentials, Comput. Phys. Commun. **196**, 36 (2015).

[77] G. Prandini, A. Marrazzo, I. E. Castelli, N. Mounet, and N. Marzari, Precision and efficiency in solid-state pseudopotential calculations, npj Comput. Mater. **4**, 72 (2018).

[78] J. P. Perdew, K. Burke, and M. Ernzerhof, Generalized gradient approximation made simple, Phys. Rev. Lett. **77**, 3865 (1996).

[79] J. P. Perdew, J. A. Chevary, S. H. Vosko, K. A. Jackson, M. R. Pederson, D. J. Singh, and C. Fiolhais, Atoms, molecules, solids, and surfaces: Applications of the generalized gradient approximation for exchange and correlation, Phys. Rev. B **46**, 6671 (1992).

[80] T. Thonhauser, S. Zuluaga, C. A. Arter, K. Berland, E. Schröder, and P. Hyldgaard, Spin signature of nonlocal correlation binding in metal-organic frameworks, Phys. Rev. Lett. **115**, 136402 (2015).

[81] S. Baroni, S. de Gironcoli, A. Dal Corso, and P. Giannozzi, Phonons and related crystal properties from density-functional perturbation theory, Rev. Mod. Phys. **73**, 515 (2001).

[82] M. Kawamura, Y. Gohda, and S. Tsuneyuki, Improved tetrahedron method for the brillouin-zone integration applicable to response functions, Phys. Rev. B **89**, 094515 (2014).

[83] R.-J. Slager, A. Mesaros, V. Juričić, and J. Zaanen, The space group classification of topological band-insulators, Nat. Phys. **9**, 98 (2013).

[84] H. C. Po, A. Vishwanath, and H. Watanabe, Symmetry-based indicators of band topology in the 230 space groups, Nat. Commun. **8**, 50 (2017).

[85] J. Kruthoff, J. de Boer, J. van Wezel, C. L. Kane, and R.-J. Slager, Topological classification of crystalline insulators through band structure combinatorics, Phys. Rev. X **7**, 041069 (2017).

[86] A. Matsugatani, S. Ono, Y. Nomura, and H. Watanabe, qeirreps: An open-source program for Quantum ESPRESSO to compute irreducible representations of bloch wavefunctions, Comput. Phys. Commun. **264**, 107948 (2021).

[87] A. A. Mostofi, J. R. Yates, Y.-S. Lee, I. Souza, D. Vanderbilt, and N. Marzari, wannier90: A tool for obtaining maximally-localised wannier functions, Comput. Phys. Commun. **178**, 685 (2008).

[88] Q. S. Wu, S. N. Zhang, H. F. Song, M. Troyer, and A. A. Soluyanov, WannierTools: An open-source software package for novel topological materials, Comput. Phys. Commun. **224**, 405 (2018).

[89] J. C. Gao, Q. S. Wu, C. Persson, and Z. J. Wang, Irvsp: To obtain irreducible representations of electronic states in the VASP, Comput. Phys. Commun. **261**, 107760 (2021).

[90] A. Sanna, C. Pellegrini, and E. K. U. Gross, Combining Eliashberg theory with density functional theory for the accurate prediction of superconducting transition temperatures and gap functions, Phys. Rev. Lett. **125**, 057001 (2020).

[91] M. Kawamura, Y. Hizume, and T. Ozaki, Benchmark of density functional theory for superconductors in elemental materials, Phys. Rev. B **101**, 134511 (2020).

[92] J. A. Flores-Livas, and A. Sanna, Superconductivity in intercalated group-IV honeycomb structures, Phys. Rev. B **91**, 054508 (2015).

[93] M. A. L. Marques, M. Lüders, N. N. Lathiotakis, G. Profeta, A. Floris, L. Fast, A. Continenza, E. K. U. Gross, and S. Massidda, Ab initio theory of superconductivity. II. Application to elemental metals, Phys. Rev. B **72**, 024546 (2005).

[94] M. Lüders, M. A. L. Marques, N. N. Lathiotakis, A. Floris, G. Profeta, L. Fast, A. Continenza, S. Massidda, and E. K. U. Gross, Ab initio theory of superconductivity. I. Density functional formalism and approximate functionals, Phys. Rev. B **72**, 024545 (2005).

[95] L. N. Oliveira, E. K. U. Gross, and W. Kohn, Density-functional theory for superconductors, Phys. Rev. Lett. **60**, 2430 (1988).

[96] G. Kresse, and J. Furthmüller, Efficient iterative schemes for ab initio total-energy calculations using a plane-wave basis set, Phys. Rev. B **54**, 11169 (1996).

[97] V. Wang, N. Xu, J.-C. Liu, G. Tang, and W.-T. Geng, VASPKIT: A user-friendly interface facilitating high-throughput computing and analysis using VASP code, Comput. Phys. Commun. **267**, 108033 (2021).

[98] M. Kawamura, R. Akashi, and S. Tsuneyuki, Anisotropic superconducting gaps in $YNi_2B_2C$: A first-principles investigation, Phys. Rev. B **95**, 054506 (2017).

[99] M. Kawamura, FermiSurfer: Fermi-surface viewer providing multiple representation schemes, Comput. Phys. Commun. **239**, 197 (2019).

[100] See Supplemental Material at http://link.aps.org/supplemental/10.1103/PhysRevB.111.144511 for the mechanical properties, Table SI, and Figs. S1–S14, which include Refs. [115-117].

[101] A. Lau, T. Hyart, C. Autieri, A. Chen, and D. I. Pikulin, Designing three-dimensional flat bands in nodal-line semimetals, Phy. Rev. X **11**, 031017 (2021).

[102] S. Carr, C. Y. Li, Z. Y. Zhu, E. Kaxiras, S. Sachdev, and A. Kruchkov, Ultraheavy and ultrarelativistic dirac quasiparticles in sandwiched graphenes, Nano Lett. **20**, 3030 (2020).

[103] J. T. Fu, J. Xu, J. Lin, J. Köhler, and S. Q. Deng, "Flat/steep band model" for superconductors containing Bi square nets, Z. Naturforsch. B **75**, 183 (2020).

[104] S. Q. Deng, C. Felser, and J. Köhler, A reverse approach to superconductivity, J. Mod. Phys. **4**, 10 (2013).

[105] J. Guo, H. Wang, F. von Rohr, Z. Wang, S. Cai, Y. Zhou, K. Yang, A. Li, S. Jiang, Q. Wu, R. J. Cava and L. Sun, Robust zero resistance in a superconducting high-entropy alloy at pressures up to 190 GPa, Proc. Natl. Acad. Sci. U.S.A. **114**, 13144 (2017).

[106] A. Majumdar, D. VanGennep, J. Brisbois, D. Chareev, A. V. Sadakov, A. S. Usoltsev, M. Mito, A. V. Silhanek, T. Sarkar, A. Hassan, O. Karis, R. Ahuja and M. Abdel-Hafiez, Interplay of charge density wave and multiband superconductivity in layered quasi-two-dimensional materials: The case of 2H-$NbS_2$ and 2H-$NbSe_2$, Phys. Rev. Mater. **4**, 084005 (2020).

[107] H. Sun, M. Huo, X. Hu, J. Li, Z. Liu, Y. Han, L. Tang, Z. Mao, P. Yang, B. Wang, J. Cheng, D.-X. Yao, G.-M. Zhang and M. Wang, Signatures of superconductivity near 80 K in a nickelate under high pressure, Nature **621**, 493 (2023).

[108] J. Karpinski, N. D. Zhigadlo, S. Katrych, K. Rogacki, B. Batlogg, M. Tortello, and R. Puzniak, $MgB_2$ single crystals substituted with Li and with Li-C: Structural and superconducting properties, Phys. Rev. B **77**, 214507 (2008).

[109] F. Tang, S. Ono, X. G. Wan, and H. Watanabe, High-throughput investigations of topological and nodal superconductors, Phys. Rev. Lett. **129**, 027001 (2022).

[110] S. Ono, and K. Shiozaki, Symmetry-based approach to superconducting nodes: Unification of compatibility conditions and gapless point classifications, Phys. Rev. X **12**, 011021 (2022).

[111] S.-J. Huang, and Y.-T. Hsu, Faithful derivation of symmetry indicators: A case study for topological superconductors with time-reversal and inversion symmetries, Phys. Rev. Res. **3**, 013243 (2021).

[112] S. Ono, H. C. Po, and H. Watanabe, Refined symmetry indicators for topological superconductors in all space groups, Sci. Adv. **6**, eaaz8367 (2020).

[113] S. Ono, H. Watanabe, F. Tang, and X. G. Wan, Topological Supercon, (2021), http://toposupercon.t.u-tokyo.ac.jp/tms/.

[114] Bilbao Crystallographic Server, https://www.cryst.ehu.es/.

[115] F. Mouhat, and F.-X. Coudert, Necessary and sufficient elastic stability conditions in various crystal systems, Phys. Rev. B **90**, 224104 (2014).

[116] S. F. Pugh, XCII. Relations between the elastic moduli and the plastic properties of polycrystalline pure metals, Philos. Mag. **45**, 823 (1954).

[117] G. Murtaza, S. K. Gupta, T. Seddik, R. Khenata, Z. A. Alahmed, R. Ahmed, H. Khachai, P. K. Jha, and S. Bin Omran, Structural, electronic, optical and thermodynamic properties of cubic $REGa_3$ (RE = Sc or Lu) compounds: ab initio study, J. Alloys Comp. **597**, 36 (2014).